\documentclass{PoS}
 \newcommand\beq{\begin{equation}}
 
 \newcommand\eeq{\end{equation}}                                               
 \newcommand\beqn{\begin{eqnarray}}
 \newcommand\eeqn{\end{eqnarray}}
 \newcommand\GeV{{\rm GeV}}
 \def\BA{\begin{eqnarray}}
 \def\BE{\begin{equation}}
 \def\BF{\begin{figure}[htb]}
 \def\BT{\begin{table}[htb]}
 \def\EA{\end{eqnarray}}
 \def\EE{\end{equation}}
 \def\EF{\end{figure}}
 \def\ET{\end{table}}

 \def\la{\langle}
 \def\ra{\rangle}
 \def\mb{\,\mbox{mb}}
 \def\fm{\,\mbox{fm}}
 \def\GeV{\,\mbox{GeV}}

 \def\lsim{\mathrel{\rlap{\lower4pt\hbox{\hskip1pt$\sim$}}
     \raise1pt\hbox{$<$}}}         %less than or approx. symbol
 \def\gsim{\mathrel{\rlap{\lower4pt\hbox{\hskip1pt$\sim$}}
     \raise1pt\hbox{$>$}}}         %greater than or approx. symbol
\title{Nuclear shadowing in the light-cone dipole approach}
                                                                                
\ShortTitle{Nuclear shadowing in the light-cone dipole approach}
                                                                                
\author{\speaker{Jan Nemchik}%
\\
         IEP SAS, Slovakia and Tech. U., Prague, Czech Republic\\
        E-mail: \email{nemcik@saske.sk}}
                                                                                
\abstract{
We study nuclear shadowing at small
Bjorken $x_{Bj}\lsim 0.01$ in the
color dipole approach. Such
a light-cone 
quantum-chromodynamics
formalism 
based on the Green function
technique 
incorporates naturally color transparency and
coherence length effects. 
The nuclear
shadowing for the $\bar{q}q$
Fock component of the photon 
is calculated using
exact numerical solution of the evolution
equation for the Green function.
At $x_{Bj}\le 0.01$ we demonstrate 
that a contribution of higher Fock 
states containing gluons
to overall nuclear shadowing becomes effective.
Numerical results for nuclear shadowing
are compared with available data from the E665 and NMC
collaborations.
Model calculations are finally tested 
with the results obtained from other models.}

\FullConference{High-pT physics at LHC\\
                 March 23-27, 2007 \\
                 University of Jyvaskyla, Jyvaskyla, Finland}
                                                                                

\begin{document}

%%%%%%%%%%%%%%%%%%%%%%%%%%%%%%%%%
\section{Introduction} 
\label{intro}
%%%%%%%%%%%%%%%%%%%%%%%%%%%%%%%%%

Nuclear shadowing (NS) in deep-inelastic scattering
(DIS) off nuclei is usually studied via 
nuclear structure functions. 
Assuming shadowing region of small Bjorken $x_{Bj}\lsim 0.01$
the structure function $F_2$ per nucleon turns out to be smaller
in nuclei than in a free nucleon.
This fact affects then the corresponding study of nuclear
effects mainly in connection with the interpretation
of the results from hadron-nucleus and heavy ion experiments.

There are different treatments of
NS depending
on the reference frame.
In the infinite momentum frame of the nucleus
NS can be interpreted
as a result of parton fusion
\cite{infinite,mq-86}
leading to a reduction of the parton density
at low Bjorken $x_{Bj}$. 
In the rest frame of the nucleus, however,
this phenomenon looks like
NS of the virtual photon hadronic fluctuations
and is occurred due to their multiple scattering
inside the target 
(see refs.~\cite{nz-91,krt-00,n-03} and references therein,
for example).

In the rest frame of the nucleus,
the dynamics
of NS is controlled by the
time scale known as the
effect of quantum coherence.
It results 
from destructive interference of
the amplitudes for which the interaction takes place on different bound
nucleons. 
It can be estimated by relying on the uncertainty principle and
Lorentz time dilation.
For the lowest Fock component
of the photon it reads,
%
%%%%%%%%%%%%%%%%%%%%%%%%%%%%%%%%%%%
 \beq
t_c = \frac{2\,\nu}{Q^2 + M_{\bar{q}q}^2}\ ,
%----------
\label{10}
%----------
 \eeq 
%%%%%%%%%%%%%%%%%%%%%%%%%%%%%%%%%%%
%
where $\nu$ and $Q^2$ is the photon energy and virtuality,
respectively
and $M_{\bar{q}q}$ is the effective mass of the ${\bar{q}q}$ pair.
It is usually called coherence time, but we also will use the term
coherence length (CL), since light-cone (LC) kinematics is assumed, $l_c=t_c$.
CL is related to the longitudinal momentum transfer $q_c=1/l_c$.

Note, that higher Fock states containing gluons,
$|\bar{q}qG\ra$, $|\bar{q}q2G\ra$, ... 
have larger effective masses than $\bar{q}q$ fluctuation
resulting in shorter coherence times. 

The Green function approach \cite{kst2}
is a very effective tool for the study
of NS
naturally incorporating the effects of CL.
Such a formalism
is based on the solution of 
the evolution equation for the Green function.
Usually, one has been forced to obtain
an analytical solution leading to
a harmonic
oscillator form (see Eq.~(\ref{142})).
It requires, however, to implement
several approximations into a rigorous quantum-mechanical
approach
like a constant nuclear density function (\ref{270}) and
a specific quadratic form (\ref{260}) of the dipole cross section.
	
We remove such approximations for the $|\bar{q}q\ra$ Fock state
solving
the evolution equation 
numerically using algorithm from
ref.~\cite{n-03}.
It is supported also by the fact that
the predictions for NS \cite{n-03}
corresponding to the lowest Fock component of the photon
showed quite large differences from
approximate calculations \cite{krt-98,krt-00}
using the harmonic oscillator Green function approach 
for $l_c \lsim R_A$
($R_A$ is the nuclear radius).

Calculations of NS
presented so far in the color dipole approach
\cite{krt-98,krt-00,n-03} 
were performed assuming only
$\bar{q}q$ fluctuations of the photon and neglecting
higher Fock components containing gluons
and sea quarks.
The effects of higher Fock states 
are included in
energy dependence of the dipole cross section, 
$\sigma_{\bar{q}q}(\vec{r},s)$\footnote{Here $\vec r$
represents the transverse separation of the $\bar{q}q$ photon
fluctuation
and $s$ is the center of mass energy squared.}.
However, investigating nuclear effects,
these Fock states lead also to
gluon shadowing (GS), which has been neglected so far
when the model predictions were compared with experimental
data.

It was demonstrated in ref.~\cite{kst2} that a contribution of 
GS to overall NS
becomes effective at small $x_{Bj}\lsim 0.01$.
It confirms the necessity for inclusion of GS
in the model predictions in the shadowing region
$\sim 0.0001\lsim x_{Bj}\lsim 0.01$, where
the data exist.
%
%%%%%%%%%%%%%%%%%%%%%%%%%%%%%%%%%%%%%%%%%%%%%%%%%%%%%%%%%%%%%%%%%%%%%%%%%%%
\section{Light-cone dipole approach to nuclear
shadowing}
\label{lc}
%%%%%%%%%%%%%%%%%%%%%%%%%%%%%%%%%%%%%%%%%%%%%%%%%%%%%%%%%%%%%%%%%%%%%%%%%%%
%

In the rest frame of the nucleus NS in the
total virtual photoabsorption cross section 
$\sigma_{tot}^{\gamma^*A}(x_{Bj},Q^2)$
(or in the structure function $F_2^A(x_{Bj},Q^2)$)
can be decomposed over different Fock components of the
virtual photon.
Then the total photoabsorption cross section on a nucleus can be formally
represented in the form
%
%%%%%%%%%%%%%%%%%%%%%%%%%%%%%%%%%%%%%%%%%%%%%%%%%%%%%%%%%%%
 \BE
\sigma_{tot}^{\gamma^*A}(x_{Bj},Q^2) =
A~\sigma_{tot}^{\gamma^*N}(x_{Bj},Q^2) -
\Delta\sigma_{tot}(x_{Bj},Q^2)\, ,
%----------
\label{110}
%----------
 \EE
%%%%%%%%%%%%%%%%%%%%%%%%%%%%%%%%%%%%%%%%%%%%%%%%%%%%%%%%%%%
%
where
$\Delta\sigma_{tot}(x_{Bj},Q^2) =
\Delta\sigma_{tot}({\bar{q}q}) +
\Delta\sigma_{tot}(\bar{q}qG) +
\Delta\sigma_{tot}(\bar{q}q2G) + ...$,
$x_{Bj}$ is the Bjorken variable and
$\sigma_{tot}^{\gamma^*N}(x_{Bj},Q^2)$ is 
total photoabsorption cross section on a nucleon 
%
%%%%%%%%%%%%%%%%%%%%%%%%%%%%%%%%%%%%%%%%%%%%%%%%%%%%%%%%%%%
 \BE
\sigma_{tot}^{\gamma^*N}(x_{Bj},Q^2) =
\int d^2 r \int_{0}^{1} d\alpha\,\Bigl
| \Psi_{\bar{q}q}(\vec{r},\alpha,Q^2)\,\Bigr |^2
~\sigma_{\bar{q}q}(\vec{r},s)\, .
%----------
\label{120}
%----------
 \EE
%%%%%%%%%%%%%%%%%%%%%%%%%%%%%%%%%%%%%%%%%%%%%%%%%%%%%%%%%%%
%

The dipole cross section $\sigma_{\bar{q}q}(\vec r,s)$ 
in Eq.~(\ref{120})
represents the interaction of a
$\bar{q}q$ dipole of transverse separation $\vec r$ with a nucleon
\cite{zkl}. It depends on the c.m. energy squared $s$. 
The dependence of $\sigma_{\bar{q}q}(\vec r,s)$ on $\vec{r}$ and energy
is universal and flavor independent
allowing to describe 
various high energy processes
in an uniform way.
$\sigma_{\bar{q}q}(\vec r,s)$ is known to vanish quadratically
$\propto r^2$ as $r\rightarrow 0$ due to color
screening (property of color transparency \cite{zkl,bbgg}).
It cannot be predicted
reliably because of poorly known higher order 
perturbative QCD (pQCD) corrections and
nonperturbative effects. 
However, $\sigma_{\bar{q}q}(\vec r,s)$ can be extracted from experimental data on
DIS and structure functions using 
reasonable parametrizations. In this case pQCD corrections
and nonperturbative effects are naturally included.

There are two popular parametrizations of $\sigma_{\bar{q}q}(\vec 
r,s)$, GBW presented in \cite{gbw} and KST suggested in \cite{kst2}.
Detailed discussion and comparison of these two parametrizations  
can be found in refs.~\cite{knst-01,n-03}.
Whereas GBW parametrization can not
be applied in the nonperturbative region of small $Q^2$, 
KST parametrization presents a good description of the 
transition down to limit of real photoproduction, $Q^2=0$. 
Because we study the shadowing region of small $x_{Bj}\lsim 0.01$
and available experimental data from the E665 and NMC collaborations cover
small and moderate values of $Q^2\lsim 2\div 3$\,GeV$^2$ we
prefer the latter parametrization. 

The KST parametrization
\cite{kst2} has the following form 
containing an explicit energy dependence,
%
%%%%%%%%%%%%%%%%%%%%%%%%%%%%%%%%
\beq
\sigma_{\bar{q}q}(r,s) = \sigma_0(s)\,\left [1 - 
exp\left ( - \frac{r^2}{R_0^2(s)}\right )\right ]\, ,
%-----------
\label{kst-1}
%-----------
\eeq
%%%%%%%%%%%%%%%%%%%%%%%%%%%%%%%%
%
where 
%
%%%%%%%%%%%%%%%%%%%%%%%%%%%%%%%%%
\beq
\sigma_0(s) = \sigma_{tot}^{\pi\,p}(s)\,\left 
(1 + \frac{3\,R_0^2(s)}{8\,\la r_{ch}^2\ra_{\pi}}\right )\, .
%------------
\label{kst-2}
%------------
\eeq
%%%%%%%%%%%%%%%%%%%%%%%%%%%%%%%%%
%

In Eq.~(\ref{kst-1}) the energy-dependent radius 
$R_0(s) = 0.88\,(s/s_0)^{-\lambda/2}\fm$ 
with $\lambda = 0.28$ and $s_0 = 1000\GeV^2$.
In Eq.~(\ref{kst-2}) 
$\sigma_{tot}^{\pi\,p}(s) = 23.6\,(s/s_0)^{0.079} + 
1.432\,(s/s_0)^{-0.45}\mb$ corresponding to the Pomeron and Reggeon parts  of the
$\pi p$ total cross section \cite{rpp-96} and 
$\la r_{ch}^2\ra_{\pi} = 0.44\fm^2$
represents the mean pion charge radius squared.

Several comments about KST parametrization are in order:

{\bf i)}
The KST dipole cross section (\ref{kst-1}) is proportional to $r^2$
for $r\rightarrow 0$, but flattens off at large $r$. 

{\bf ii)}
The saturation scale $R_0(s)$ decreases with increasing $s$. 
The energy dependence of $\sigma_{\bar{q}q}(r,s)$
correlates with $r$. At small $r$ the dipole cross section rises with a 
hard pomeron intercept, $0.36$, and at large separations it still depends
with a soft pomeron intercept, $0.08$, on energy.

{\bf iii)}
In contrast to GBW parametrization \cite{gbw}, energy dependent ansatz,
Eq.~(\ref{kst-1}), allows to describe simultaneously 
hadron-hadron scattering and DIS. 
This improvement at large separations leads to a worse description
of the short-distance part of the dipole cross section which is 
responsible for the behavior of $F_2^p(x_{Bj},Q^2)$ at large $Q^2$.
To satisfy Bjorken scaling the dipole cross section at small $r$
must be a function of the product $s\,r$ which is not the case
for the parametrization in Eq.~(\ref{kst-1}).

{\bf iv)}
The form of Eq.~(\ref{kst-1})
successfully describes the data for DIS at small $x_{Bj}$ only up to
$Q^2\approx 10\GeV^2$. Nevertheless, this interval of $Q^2$ is
sufficient for the purpose of the present paper. \\

The second ingredient,
$\Psi_{\bar{q}q}({\vec{r}},\alpha,Q^2)$,
in (\ref{120}) is
the perturbative distribution amplitude (``wave function'') of the $\bar{q}q$ 
Fock component of the photon.
It depends on the
photon virtuality $Q^2$ and the relative share $\alpha$ of the photon
momentum carried by the quark. 
The corresponding form
for transversely (T) and longitudinally (L) polarized photons 
can be found in refs.~\cite{lc,bks-71,nz-91}:
%
%%%%%%%%%%%%%%%%%%%%%%%%%%%%%%%%%%%%%%%%%%%%%%
 \BE
\Psi_{\bar{q}q}^{T,L}({\vec{r}},\alpha,Q^2) =
\frac{\sqrt{N_{C}\,\alpha_{em}}}{2\,\pi}\,\,
Z_{q}\,\bar{\chi}\,\hat{O}^{T,L}\,\chi\, 
K_{0}(\epsilon\,r)
%----------
\label{122}
%----------
 \EE
%%%%%%%%%%%%%%%%%%%%%%%%%%%%%%%%%%%%%%%%%%%%%%
%
where $\chi$ and $\bar{\chi}$ are the spinors of the quark and
antiquark, respectively; $Z_{q}$ is the quark charge,
$N_{C} = 3$ is the number of colors. 
$K_{0}(\epsilon r)$ is a modified Bessel
function with 
%
%%%%%%%%%%%%%%%%%%%%%%%%%%%%%%%%%%%%%%%%%%%%
 \BE
\epsilon^{2} =
\alpha\,(1-\alpha)\,Q^{2} + m_{q}^{2}\ ,
%----------
\label{123}
%----------
 \EE
%%%%%%%%%%%%%%%%%%%%%%%%%%%%%%%%%%%%%%%%%%%%
%
where $m_{q}$ is the quark mass.
The form of the operators $\hat{O}^{T,L}$ can be found
in \cite{kst2}.

The distribution amplitude Eq.~(\ref{122}) controls 
the transverse $\bar{q}q$ separation 
with the mean value,
%
%%%%%%%%%%%%%%%%%%%%%%%%%%%%%%%%%%%%%%%%%%%%%%%%%%%%%%%%
 \BE
\la r\ra \sim \frac{1}{\epsilon} = 
\frac{1}{\sqrt{Q^{2}\,\alpha\,(1-\alpha) + m_{q}^{2}}}\,.
\label{130}
 \EE
%%%%%%%%%%%%%%%%%%%%%%%%%%%%%%%%%%%%%%%%%%%%%%%%%%%%%%%%
%

 For very asymmetric $\bar{q}q$ pairs with $\alpha$ or $(1-\alpha) \lsim
m_q^2/Q^2$ the mean transverse separation $\la r\ra \sim 1/m_q$ becomes
huge since one must use current quark masses within pQCD. A popular
recipe to fix this problem is to introduce an effective quark mass
$m_{eff}\sim \Lambda_{QCD}$ which should represent the nonperturbative
interaction effects between $q$ and $\bar{q}$. 
However, we introduce this interaction explicitly
using corresponding phenomenology based on the LC Green function
approach developed in \cite{kst2}.

The Green function $G_{\bar{q}q}(\vec{r_2},z_2;\vec{r_1},z_1)$ 
describes the 
propagation of an interacting $\bar{q}q$ pair between points with
longitudinal coordinates $z_1$ and $z_2$ and with initial and final
separations $\vec{r_1}$ and $\vec{r_2}$. This
Green function satisfies the 
two-dimensional Schr\"odinger equation, 
%
%%%%%%%%%%%%%%%%%%%%%%%%%%%%%%%%%%%%%%%%%%%%%%%%%%%%%%%%%%%%%%%%%%%%%%%%
 \BE
i\frac{d}{dz_2}\,G_{\bar{q}q}(\vec{r_2},z_2;\vec{r_1},z_1)=
\left[\frac{\epsilon^{2} - \Delta_{r_{2}}}{2\,\nu\,\alpha\,(1-\alpha)}
+V_{\bar{q}q}(z_2,\vec{r_2},\alpha)\right]
G_{\bar{q}q}(\vec{r_2},z_2;\vec{r_1},z_1)\ ,
\label{135}  
 \EE
%%%%%%%%%%%%%%%%%%%%%%%%%%%%%%%%%%%%%%%%%%%%%%%%%%%%%%%%%%%%%%%%%%%%%%%%
%
with the boundary condition 
%
%%%%%%%%%%%%%%%%%%%%%%%%%%%%%%%%%%%%%%%%%%%%%%%%%%%%%%%%%%%%%%%%%%%%%%%%
 \BE
G_{\bar{q}q}(\vec{r_2},z_2;\vec{r_1},z_1)|_{z_2=z_1}=
\delta^2(\vec{r_1}-\vec{r_2})\, .
\label{136}  
 \EE
%%%%%%%%%%%%%%%%%%%%%%%%%%%%%%%%%%%%%%%%%%%%%%%%%%%%%%%%%%%%%%%%%%%%%%%%
%
In Eq.~(\ref{135}) $\nu$ is the photon energy and the Laplacian 
$\Delta_{r}$ acts on the coordinate $r$.  

Studying propagation of a $\bar{q}q$ in vacuum
the LC potential $V_{\bar{q}q}(z_2,\vec{r_2},\alpha)$ in (\ref{135}) 
contains only the real part, which
is responsible for
the interaction between the $q$ and $\bar{q}$.  
For the oscillator form 
of this potential,
%
%%%%%%%%%%%%%%%%%%%%%%%%%%%%%%%%%%%%%%%%%%%%%%%
 \BE  
{\rm Re}\,V_{\bar{q}q}(z_2,\vec{r_2},\alpha) =
\frac{a^4(\alpha)\,\vec{r_2}\,^2} 
{2\,\nu\,\alpha(1-\alpha)}\ ,
\label{140} 
 \EE
%%%%%%%%%%%%%%%%%%%%%%%%%%%%%%%%%%%%%%%%%%%%%%%
%
one can obtain the analytical solution
of the evolution equation (\ref{135}) represented by the
harmonic oscillator Green function \cite{fg},
%
%%%%%%%%%%%%%%%%%%%%%%%%%%%%%%%%%%%%%%%%%%%%%%%%%%%%%%%%%%
 \BA 
G_{\bar{q}q}(\vec{r_2},z_2;\vec{r_1},z_1) =
\frac{a^2(\alpha)}{2\;\pi\;i\;
{\rm sin}(\omega\,\Delta z)}\, {\rm exp}
\left\{\frac{i\,a^2(\alpha)}{{\rm sin}(\omega\,\Delta z)}\,
\Bigl[(r_1^2+r_2^2)\,{\rm cos}(\omega \;\Delta z) -
2\;\vec{r_1}\cdot\vec{r_2}\Bigr]\right\}
\nonumber\\ \times {\rm exp}\left[- 
\frac{i\,\epsilon^{2}\,\Delta z}
{2\,\nu\,\alpha\,(1-\alpha)}\right] \ , 
\label{142} 
 \EA
%%%%%%%%%%%%%%%%%%%%%%%%%%%%%%%%%%%%%%%%%%%%%%%%%%%%%%%%%%%%
%
where $\Delta z=z_2-z_1$ and 
%
%%%%%%%%%%%%%%%%%%%%%%%%%%%%%%%%%%%%%%%%%%%%%%%%%%%%%%%%%%%%
 \BE \omega = \frac{a^2(\alpha)}{\nu\;\alpha(1-\alpha)}\ .
\label{144} 
 \EE
%%%%%%%%%%%%%%%%%%%%%%%%%%%%%%%%%%%%%%%%%%%%%%%%%%%%%%%%%%%%
%
The shape of the function $a(\alpha)$ in Eq.~(\ref{140}) is
presented and discussed in ref.~\cite{kst2}.

Matrix element (\ref{120}) contains the LC wave function squared,
which has the following form for T and L polarizations in
the limit of vanishing interaction between $\bar{q}$ and $q$,
%
%%%%%%%%%%%%%%%%%%%%%%%%%%%%%%%%%%%%%%%%%%%%%%%%%%%%%%%%%%%
 \BE
\Bigl |\Psi^{T}_{\bar{q}q}(\vec r,\alpha,Q^2)\,\Bigr |^2 =
\frac{2\,N_C\,\alpha_{em}}{(2\pi)^2}\,
\sum_{f=1}^{N_f}\,Z_f^2
\left[m_f^2\,K_0(\epsilon,r)^2
+ [\alpha^2+(1-\alpha)^2]\,\epsilon^2\,K_1(\epsilon\,r)^2\right]\ ,
%----------
\label{197a}
%----------
 \EE
%%%%%%%%%%%%%%%%%%%%%%%%%%%%%%%%%%%%%%%%%%%%%%%%%%%%%%%%%%%
%
%and
%
%%%%%%%%%%%%%%%%%%%%%%%%%%%%%%%%%%%%%%%%%%%%%%%%%%%%%%%%%%%
 \BE
\Bigl |\Psi^{L}_{\bar{q}q}(\vec r,\alpha,Q^2)\,\Bigr |^2 =
\frac{8\,N_C\,\alpha_{em}}{(2\pi)^2}\,
\sum_{f=1}^{N_f}\,Z_f^2
\,Q^2\,\alpha^2(1-\alpha)^2\,
K_0(\epsilon\,r)^2\ ,
%----------
\label{197b}
%----------
 \EE
%%%%%%%%%%%%%%%%%%%%%%%%%%%%%%%%%%%%%%%%%%%%%%%%%%%%%%%%%%%
%   
where $K_0$ and $K_1$ are the modified Bessel functions.

The effects of higher Fock states 
$|\bar{q}q\ra$, $|\bar qqG\ra$, 
$|\bar{q}q2G\ra$, ...
are implicitly incorporated
into the energy 
dependence of the dipole cross section
$\sigma_{\bar{q}q}(\vec{r},s)$ 
naturally included in realistic KST parametrization
Eq.~(\ref{kst-1}). 

Investigating
DIS on nuclear targets, the corresponding 
formula for NS,
$\Delta\sigma_{tot}(x_{Bj},Q^2) =
\Delta\sigma_{tot}(\bar{q}q)$,
representing the shadowing correction for the lowest
$\bar{q}q$ Fock state,
has the following form \cite{krt-98}
%
%%%%%%%%%%%%%%%%%%%%%%%%%%%%%%%%%%%%%%%%%%%%%%%%%%%%%%%%%%%
 \BE
\Delta\sigma_{tot}(x_{Bj},Q^2) = \frac{1}{2}~{Re}~\int d^2 b \int_{-\infty}^{
\infty} dz_1 ~\rho_{A}(b,z_1) \int_{z_1}^{\infty} dz_2~
\rho_A(b,z_2) 
\int_{0}^{1} d\alpha ~A(z_1,z_2,\alpha)\, ,
%----------
\label{230}
%----------
 \EE
%%%%%%%%%%%%%%%%%%%%%%%%%%%%%%%%%%%%%%%%%%%%%%%%%%%%%%%%%%%
%
with 
%
%%%%%%%%%%%%%%%%%%%%%%%%%%%%%%%%%%%%%%%%%%%%%%%%%%%%%%%%%%%
 \BE
A(z_1,z_2,\alpha) 
= \int d^2 r_2 ~\Psi^{*}_{\bar{q}q}(\vec{r_2},\alpha,Q^2)
~\sigma_{\bar{q}q}(r_2,s) \int d^2 r_1
~G_{\bar{q}q}(\vec{r_2},z_2;\vec{r_1},z_1)
~\sigma_{\bar{q}q}(r_1,s)
~\Psi_{\bar{q}q}(\vec{r_1},\alpha,Q^2)\, .
%----------
\label{240}
%----------
 \EE
%%%%%%%%%%%%%%%%%%%%%%%%%%%%%%%%%%%%%%%%%%%%%%%%%%%%%%%%%%%
%
In Eq.~(\ref{230}) 
$\rho_{A}({b},z)$ represents the nuclear density function defined
at the point with longitudinal coordinate $z$ and impact
parameter $\vec{b}$. 
%
%****************************************************************
%************************ FIG.1 *********************************
%****************************************************************
 \begin{figure}[tbh]
\includegraphics{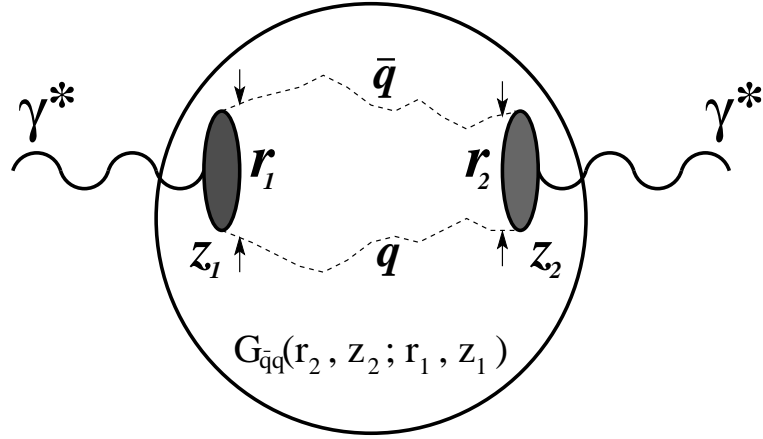}
\begin{center}
\vspace{6.7cm}
\parbox{13cm} 
{\caption[Delta]
 {A cartoon for the shadowing term $\Delta\sigma_{tot}(x_{Bj},Q^2)
= \Delta\sigma_{tot}(\bar{q}q)$ in (\ref{230}). Propagation of the
$\bar{q}q$ pair through the nucleus is described by the Green
function $G_{\bar{q}q}(\vec{r_2},z_2;\vec{r_1},z_1)$, which results
from the summation over different paths of the $\bar{q}q$ pair.}
%%%%%%%%%%%%%%%%%%%%%%%%%
 \label{shad}}
%%%%%%%%%%%%%%%%%%%%%%%%%
\end{center}
 \end{figure}
%****************************************************************
%

The shadowing term $\Delta\sigma_{tot}(x_{Bj},Q^2) =
\Delta\sigma_{tot}(\bar{q}q)$ in (\ref{230}) is illustrated in
Fig.~\ref{shad}. At the point $z_1$ the initial photon diffractively
produces the $\bar{q}q$ pair ($\gamma^*N\to \bar{q}qN$) with
transverse separation $\vec{r_1}$. The $\bar{q}q$ pair then
propagates through the nucleus along arbitrary curved trajectories,
which are summed over, and arrives at the point $z_2$ with
transverse separation $\vec{r_2}$. The initial and final separations
are controlled by the LC wave function of the $\bar{q}q$ Fock
component of the photon $\Psi_{\bar{q}q}(\vec{r},\alpha,Q^2)$. During
propagation through the nucleus the $\bar{q}q$ pair interacts with
bound nucleons via the dipole cross section $\sigma_{\bar{q}q}(r,s)$,
which depends on the local transverse separation $\vec{r}$. The
Green function $G_{\bar{q}q}(\vec{r_2},z_2;\vec{r_1},z_1)$ describes
the propagation of the $\bar{q}q$ pair from $z_1$ to $z_2$.

Describing propagation of the $\bar{q}q$ pair in nuclear medium
the Green function $G_{\bar{q}q}(\vec{r_2},z_2;\vec{r_1},z_1)$ 
satisfies again the time-dependent
two-dimensional Schr\"odinger equation (\ref{135}).
However, the potential in this case acquires 
in addition also the imaginary part responsible for
attenuation of the $\bar{q}q$ photon fluctuation in the medium
and has the following form
%
%%%%%%%%%%%%%%%%%%%%%%%%%%%%%%%%%%%%%%%%%%%%%%%%%%%%%%%%%%%%
 \BE
Im V_{\bar{q}q}(z_2,\vec r,\alpha) = - 
\frac{\sigma_{\bar{q}q}(\vec r,s)}{2}\,\rho_{A}({b},z_2)\, .
%----------
\label{250}
%----------
 \EE
%%%%%%%%%%%%%%%%%%%%%%%%%%%%%%%%%%%%%%%%%%%%%%%%%%%%%%%%%%%%
%

As we already mentioned above
the analytical solution of Eq.~(\ref{135}) is known only for the 
harmonic oscillator potential $V_{\bar{q}q}(r)\propto r^2$, 
i.e. 
%
%%%%%%%%%%%%%%%%%%%%%%%%%%%%%%%%%%%%%%%%%%
 \beq
\sigma_{\bar{q}q}(r,s) = C(s)\,r^2\ ,
%----------
\label{260}
%----------
 \eeq
%%%%%%%%%%%%%%%%%%%%%%%%%%%%%%%%%%%%%%%%%%
%
and uniform nuclear density 
%
%%%%%%%%%%%%%%%%%%%%%%%%%%%%%%%%%%%%%%%%%%
 \beq
\rho_A(b,z) = \rho_0~\Theta(R_A^2-b^2-z^2)\, 
%----------
\label{270}
%----------
 \eeq
%%%%%%%%%%%%%%%%%%%%%%%%%%%%%%%%%%%%%%%%%%
%
in Eq.~(\ref{250}) for the imaginary part of the LC
potential.

In this case the solution of Eq.~(\ref{135}) has the same form
as Eq.~(\ref{142}), expect that one should replace $\omega\Longrightarrow
\Omega$ and $a^2(\alpha)\Longrightarrow b(\alpha)$, where 
%
%%%%%%%%%%%%%%%%%%%%%%%%%%%%%%%%%%%%%%%%%%%%%
\BE
\Omega = 
\frac{b(\alpha)}{\nu \alpha (1 - \alpha)}
=
\frac{
\sqrt{a^4(\alpha) - i\,\rho_A(b,z)\,\nu\,\alpha\,(1 - \alpha)\,C(s)}}
{\nu\,\alpha\,(1 - \alpha)}\, .
%----------
\label{280}
%----------
\EE
%%%%%%%%%%%%%%%%%%%%%%%%%%%%%%%%%%%%%%%%%%%%%
%

Determination of the energy dependent factor $C(s)$ in Eq.~(\ref{260}) 
and the mean nuclear density 
$\rho_0$ in Eq.~(\ref{270}) can be realized 
by the procedure described in \cite{krt-00,n-03}.

In the general case when there is no
restrictions for $l_c$,
if $l_c\sim R_A$ 
one has to take into account the variation of the transverse size
$r$ during propagation of the $\bar{q}q$ pair through the nucleus.
The overall total photoabsorption cross section on a nucleus
$\sigma^{\gamma^*A} = \sigma_T^{\gamma^*A} +
\epsilon'\,\sigma_L^{\gamma^*A}$, assuming that the photon
polarization $\epsilon'=1$.
If one takes into account only $\bar{q}q$ Fock component
of the photon
the full expression after summation over all flavors, colors,
helicities and spin states has the following form \cite{bgz-98}
%
%%%%%%%%%%%%%%%%%%%%%%%%%%%%%%%%%%%%%%%%%%%%%%%%%%%%%%%
\BA
\sigma^{\gamma^*A}(x_{Bj},Q^2) 
&=&
A\,\sigma^{\gamma^*N}(x_{Bj},Q^2) 
- 
\Delta\,\sigma(x_{Bj},Q^2)
\nonumber \\
&=& 
A\,\int\,d^2r\,\int_{0}^{1}\,d\alpha\,\sigma_{\bar{q}q}(r,s)
\,\Biggl (\Bigl 
|\Psi^T_{\bar{q}q}(\vec{r},\alpha,Q^2)\Bigr 
|^2 +
\Bigl |\Psi^L_{\bar{q}q}(\vec{r},\alpha,Q^2)\Bigr 
|^2\Biggr )
\nonumber \\
\,&-&\,
\frac{N_C\,\alpha_{em}}{(2\pi)^2}\sum_{f=1}^{N_f}\,Z_f^2\,\,\Re e\,
\int\,d^2b\,\int_{-\infty}^{\infty}\,dz_1\,\int_{z_1}^{\infty}\,
dz_2\,\int_{0}^{1}\,d\alpha\,\int\,d^2r_1\,\int\,d^2r_2
\nonumber\\
\times\,
\rho_A(b,z_1)&\rho_A(b,z_2)&
\sigma_{\bar{q}q}(r_2,s)
\sigma_{\bar{q}q}(r_1,s)
\,\Biggl\{\Bigl[\,\alpha^2 
+ (1 - 
\alpha)^2\,\Bigr]
\,\epsilon^2
\,\frac{\vec{r_1}\,\cdot\,\vec{r_2}}{r_1\,r_2}\,
K_1(\epsilon\,r_1)\,K_1(\epsilon\,r_2)
\nonumber\\
\,&+&\,\Bigl[\,m_f^2 + 
4\,Q^2\,\alpha^2\,(1 - 
\alpha)^2\,\Bigr]\,
K_0(\epsilon\,r_1)\,K_0(\epsilon\,r_2)\Biggr\}\,
G_{\bar{q}q}(\vec{r_2},z_2;\vec{r_1},z_1) 
\, .
%----------
\label{320}
%----------
\EA
%%%%%%%%%%%%%%%%%%%%%%%%%%%%%%%%%%%%%%%%%%%%%%%%%%%%%%% 
% 
The shape of 
$\Bigl |\,\Psi^{T,L}_{\bar{q}q}(\vec{r},\alpha,Q^2)\,\Bigr |^2$
is given by Eqs.~(\ref{197a})
and (\ref{197b}), respectively.

In the high energy limit,
$l_c\gg R_A$, 
the transverse separation $r$ between
$\bar{q}$ and $q$ does not vary 
during propagation through the nucleus (Lorentz time dilation).
Consequently, 
the kinetic
term in Eq.~(\ref{135}) can be neglected and the Green function
reads
%
%%%%%%%%%%%%%%%%%%%%%%%%%%%%%%%%%%%%%%%%%%%%%%%%%%%%%
\BA
G_{\bar{q}q}(b;\vec{r_2},z_2;\vec{r_1},z_1)|_{\nu\to\infty} = 
\delta(\vec{r_2}-\vec{r_1})\,\exp\Biggl[ - \frac{1}{2}\,
\sigma_{\bar{q}q}(r_2,s)\,\int_{z_1}^{z_2}\,dz\,\rho_A(b,z)\Biggr]\,.
%----------
\label{330}
%----------
\EA
%%%%%%%%%%%%%%%%%%%%%%%%%%%%%%%%%%%%%%%%%%%%%%%%%%%%%
%
After substitution of the expression (\ref{330}) into Eq.~(\ref{320}),
one arrives at the following results:
%
%%%%%%%%%%%%%%%%%%%%%%%%%%%%%%%%%%%%%%%%%%%%%%%%%%%%%%%
\BA
&&\sigma^{\gamma^*A}_{npt}(x_{Bj},Q^2) =
2\,\int\,d^2b\,\int\,d^2r\,\int_0^1\,d\alpha
\left\{1 - \exp\,\Bigl[ - \frac{1}{2}\,\sigma_{\bar{q}q}(r,s)\, 
T_A(b)\Bigr]\,\right\} \nonumber\\
\times\,&&\frac{2\,N_C\,\alpha_{em}}{(2\,\pi)^2}
\sum_{f=1}^{N_f}\,Z_f^2\,
\Biggl\{\Bigl[\,\alpha^2 + (1 - \alpha)^2\,\Bigr]
K_1(\epsilon\,r)^2\,
 + \,\Bigl[m_f^2 + 4\,Q^2\,\alpha^2\,(1 - \alpha)^2\Bigr]
K_0^2(\epsilon\,r)\,\Biggr\} ,
%----------
\label{335}
%----------
\EA
%%%%%%%%%%%%%%%%%%%%%%%%%%%%%%%%%%%%%%%%%%%%%%%%%%%%%%%
%
where
%
%%%%%%%%%%%%%%%%%%%%%%%%%%%
\beq
T_A(b) = \int_{-\infty}^{\infty}\,dz\,\rho_A(b,z)\,
%-----------
\label{300}
%-----------
\eeq
%%%%%%%%%%%%%%%%%%%%%%%%%%%
%
is the nuclear thickness calculated with the realistic Wood-Saxon
form of the nuclear density with parameters taken from \cite{saxon}.

At photon polarization parameter $\epsilon'=1$
the structure function ratio 
$F_2^A(x_{Bj},Q^2)/F_2^N(x_{Bj},Q^2)$ can be expressed via
ratio of the total photoabsorption cross sections
%
%%%%%%%%%%%%%%%%%%%%%%%%%%%%%%
\BA
\frac{F_2^A(x_{Bj},Q^2)}{F_2^N(x_{Bj},Q^2)} = 
\frac{\sigma_T^{\gamma^*A}(x_{Bj},Q^2)
+ \sigma_L^{\gamma^*A}(x_{Bj},Q^2)}
{\sigma_T^{\gamma^*N}(x_{Bj},Q^2)
+ \sigma_L^{\gamma^*N}(x_{Bj},Q^2)}\, ,
%----------
\label{340}
%----------
\EA
%%%%%%%%%%%%%%%%%%%%%%%%%%%%%%
%
where the numerator on right-hand side (r.h.s.) is given by 
Eq.~(\ref{320}), whereas denominator can be expressed
as the first term of Eq.~(\ref{320})
divided by the mass number $A$.

The nonperturbative $\bar{q}-q$ interaction is included
by replacements $K_0(\epsilon\,r)/2\,\pi \Longrightarrow
\Phi_0(\epsilon,r,\lambda)$ and
$K_1(\epsilon\,r)/2\,\pi \Longrightarrow
\Phi_1(\epsilon,r,\lambda)$
in all perturbative expressions.
The corresponding functions $\Phi_{0,1}$ and
parameter $\lambda$
are defined in ref.~\cite{kst2}.

We solve
the evolution equation for the Green function, Eq.~(\ref{135}), 
numerically using algorithm from ref.~\cite{n-03}.
Such an exact solution is performed 
for realistic KST parametrization of the dipole
cross section
(\ref{kst-1})
and nuclear density function
in the realistic Wood-Saxon form with parameters 
taken from ref.~\cite{saxon}.

Finally we would like to emphasize 
(see also the next section)
that the $\bar{q}q$ Fock component 
of the photon represents the higher twist shadowing correction 
\cite{krt-00}, and vanishes at large quark masses as $1/m_f^2$.
This does not happen for higher Fock states containing gluons,
which lead to GS. Therefore
GS represents the leading twist shadowing correction \cite{kst2}.
Moreover, a steep energy dependence of the dipole cross section 
$\sigma_{\bar{q}q}(r,s)$ (see Eq.~(\ref{kst-1})) 
especially at smaller dipole sizes $r$ causes
a steep energy rise of both corrections.

%
%
%
%%%%%%%%%%%%%%%%%%%%%%%%%%%%%%%%%%%%%%%%%%%%%%%%%%%%%%%%%%
\section{Gluon shadowing}\label{glue-shadow}
%%%%%%%%%%%%%%%%%%%%%%%%%%%%%%%%%%%%%%%%%%%%%%%%%%%%%%%%%%
%
%
%

Investigating nuclear effects at small $x_{Bj}\lsim 0.01$,
the higher Fock states of the photon 
containing gluons also contribute to NS.
Because of a 
shorter coherence time (lifetime) 
of these fluctuations,
GS will be dominated at higher energies, i.e. at smaller
$x_{Bj}$- values than NS coming from the
lowest $|\bar{q}q>$ Fock component.
However, no data for GS are available and
one should rely on calculations.

NS for $|\bar{q}q\ra$ Fock component
of the photon is dominated by T photon
polarizations, because the corresponding photoabsorption
cross section is scanned at larger dipole sizes than
for the L photon polarization.
The transverse $\bar{q}q$  
separation is controlled by the distribution amplitude
Eq.~(\ref{122}), with the mean value given by Eq.~(\ref{130}).
Contributions of large size dipoles come from the
asymmetric $\bar{q}q$ fluctuations of the virtual photon,
when the quark and antiquark
in the photon carry a very large ($\alpha\to 1$) and a very small 
fraction ($\alpha\to 0$) of the photon momentum, and vice versa. 
The LC wave function for L photons (\ref{197b}) 
contains a term $\alpha^2\,(1-\alpha)^2$, which makes considerably smaller the 
contribution from asymmetric $\bar{q}q$ configurations than for   
T photons (see Eq.~(\ref{197a})). Consequently, in   
contrast to T photons, all $\bar{q}q$ dipoles from
L photons have a size squared $\propto 1/Q^2$ and the
double-scattering term vanishes as $\propto 1/Q^4$.
The leading-twist contribution for the shadowing of L
photons arises from the $|\bar{q}qG\ra$ Fock component
of the photon because the gluon can propagate relatively 
far from the $\bar{q}q$ pair, although the $\bar{q}$-$q$ separation
is of the order $1/Q^2$. After radiation of the gluon the pair
is in an octet state and consequently the $|\bar{q}qG\ra$
state represents a $GG$ dipole. Then the corresponding
correction to the L cross section is just GS.  

Interpretation of GS is 
reference frame dependent.
In the infinite momentum frame this phenomenon looks analogical as
gluon-gluon fusion. 
Within a parton model interpretation,
the gluon clouds
of nucleons which have the same impact parameter overlap at small
$x_{Bj}$ in longitudinal direction. This allows gluons originated from
different nucleons to fuse leading to a gluon density which is not
proportional to the density of nucleons any more. This is GS.

In the rest frame of the nucleus phenomenon of GS
corresponds to the process of gluon radiation and shadowing corrections
related to multiple interactions of the radiated gluons in the nuclear
medium. This is a coherence phenomenon known as the
Landau-Pomeranchuk effect, namely the suppression of bremsstrahlung by
interference of radiation from different scattering centers. It demands
a sufficiently long coherence time of radiation, a condition equivalent
to demanding a small Bjorken $x_{Bj}$ in the parton model.

There are still very few numerical evaluations of GS
in the literature, all done in the rest frame of the nucleus.
GS can be identified
as the shadowing correction to the L cross
section coming from the $GG$ dipole representing
$|\bar{q}qG\ra$ Fock component of the photon.
For evaluation of GS is important to know about 
the transverse size of this $GG$ dipole. 
This size has been extracted in ref.~\cite{kst2}
from data for diffractive excitation of the incident
hadrons to the states of large mass, the so called triple-Pomeron region.
It was found that the mean
dipole size of the $GG$ system 
(radius of propagation of the LC gluons) is
rather small , $r_0\approx 0.3\,\fm$ \cite{k3p}. 
It results in a not very strong onset of GS.

The smallness of the size of quark-gluon fluctuations 
has been incorporated via
a nonperturbative LC potential in the Schr\"odinger equation
for the Green function describing the propagation of a quark-gluon
system.  
The strength of the potential was fixed by data on high mass
($M_X^2$) diffraction $pp\to pX$ \cite{kst2}. 
This approach allows to extend the methods of pQCD to the region
of small $Q^2\lsim Q_0^2 = 4/r_0^2$.
At higher $Q^2\gsim Q_0^2$ GS slowly (logarithmically) decreases in 
accordance
with expectations based on the evolution equation \cite{mq-86}.

%
%****************************************************************
%************************ FIG.1 *********************************
%****************************************************************
 \begin{figure}[htb]
\includegraphics{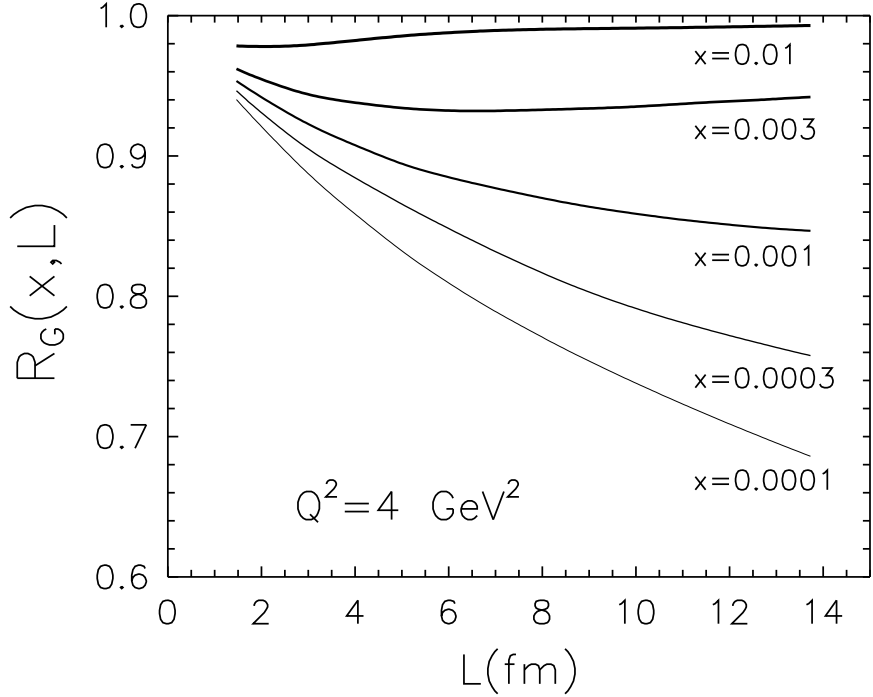}
\begin{center}
\vspace{9.2cm}
\parbox{14.0cm}
{\caption[Delta]
{
The ratio of the nucleus-to-nucleon gluon densities as function of the
thickness of the nucleus, $L=T(b)/\rho_0$, at $Q^2=4\,GeV^2$ and different fixed
values of $x_{Bj}$. Figure is taken from ref.~\cite{knst-01}.
}
%%%%%%%%%%%%%%%%%%%%%%%%%
 \label{glue-shad}}
%%%%%%%%%%%%%%%%%%%%%%%%%
\end{center}
 \end{figure}
%****************************************************************
%

We repeated the calculations \cite{kst2} of the ratio of the gluon
densities in nuclei and nucleon,
%
%%%%%%%%%%%%%%%%%%%%%%%%%%%%%%%%%%%%%%%%%%%%%%%%%%%%%%%%%%%%%%%%%%%%%%
 \beq
R_G(x_{Bj},Q^2)=\frac{G_A(x_{Bj},Q^2)}{A\,G_N(x_{Bj},Q^2)} \approx
1 - \frac{\Delta\sigma_{tot}(\bar{q}qG)}
{\sigma_{tot}^{\gamma^*A}}\ ,
\label{RG}
 \eeq
%%%%%%%%%%%%%%%%%%%%%%%%%%%%%%%%%%%%%%%%%%%%%%%%%%%%%%%%%%%%%%%%%%%%%%
%
where $\Delta\sigma_{tot}(\bar{q}qG)$ is the inelastic correction to the total
cross section $\sigma_{tot}^{\gamma^*A}$ related to the creation of a
$|\bar{q}qG\ra$ intermediate Fock state.
Further calculation details can be found in \cite{kst2}. 
As an illustration of not very strong onset of GS, here we present
$R_G(x_{Bj},Q^2)$, Eq.~(\ref{RG}), for different nuclear thicknesses
$T_A(b)$. Using an approximation of constant nuclear density (see
Eq.~(\ref{270})), $T_A(b)=\rho_0\,L$, where $L=2\,\sqrt{R_A^2 - b^2}$, the 
ratio
$R_G(x_{Bj},Q^2)$ is also implicitly a function of $L$. An example for
the calculated $L$-dependence of $R_G(x_{Bj},Q^2)$ at $Q^2=4\,\GeV^2$ is
depicted in Fig.~\ref{glue-shad} for different values of $x_{Bj}$.

We calculated GS only for the lowest Fock component
containing just one LC gluon. 
Inclusion of higher
multigluon Fock components is still a challenge. However, their effect
can be essentially taken into account by eikonalization of the calculated
$R_G(x_{Bj},Q^2)$, i.e. the dipole
cross section, which is proportional to the gluon density at small
separations, should be renormalized everywhere,
%
%%%%%%%%%%%%%%%%%%%%%%%%%%%%%%%%%%%
 \beq
\sigma_{\bar{q}q} \Rightarrow 
R_G\,\sigma_{\bar{q}q}\ .
\label{1100}
 \eeq 
%%%%%%%%%%%%%%%%%%%%%%%%%%%%%%%%%%%
%

According to Eq.~(\ref{1100})
we will demonstrate that
GS suppresses the total photoabsorption
cross section on a nucleus $\sigma_{tot}^{\gamma^* A}(x_{Bj},Q^2)$.
We expect a non-negligible effect of GS
in the shadowing region of small $x_{Bj}\lsim (0.001\div 0.01)$ 
and at small 
and medium values of $Q^2\sim 2\div 3\,$GeV$^2$
corresponding to the
kinematic range of available data.

%
%%%%%%%%%%%%%%%%%%%%%%%%%%%%%%%%%%%%%%%%%%%%%%%%%%%%%%%%%%
\section{Numerical results}
\label{results}
%%%%%%%%%%%%%%%%%%%%%%%%%%%%%%%%%%%%%%%%%%%%%%%%%%%%%%%%%%
%

Here
we present the available data vs.
realistic predictions for NS
in DIS
based on exact numerical solutions of the
evolution equation for the Green function
corresponding to the lowest $\bar{q}q$ Fock component
of the photon.
Such a comparison is performed for the shadowing
region of small $x_{Bj}\lsim 0.01$.

We take into account also a contribution of GS
which should increase overall nuclear suppression.
The effects of GS are calculated for the lowest
Fock component containing just one LC gluon.
Inclusion of higher Fock components with
more gluons is realized by eikonalization of the
calculated $R_G(x_{Bj},Q^2)$, i.e
using renormalization (\ref{1100}).

%
%****************************************************************
%************************ FIG.2 *********************************
%****************************************************************
 \begin{figure}[htb]
\includegraphics{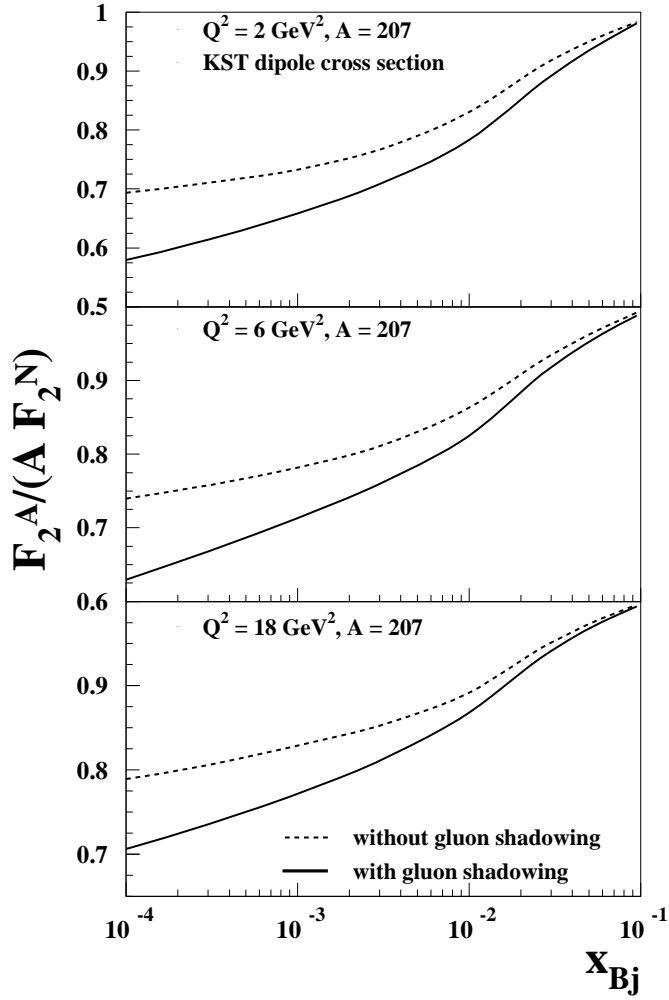}
\begin{center}
\vspace{12.9cm}
\parbox{14.0cm}
{\caption[Delta]
 {Nuclear shadowing for lead.
Calculations correspond to exact
numerical solution of the evolution equation for the Green function,
$G_{\bar{q}q}(\vec{r_2},z_2;\vec{r_1},z_1)$,
for the lowest Fock component of the photon
using KST \cite{kst2}
parametrization of the dipole cross section and
realistic nuclear density
function of the Woods-Saxon form \cite{saxon}.
The solid and dashed curves
represent the predictions
calculated with and without contribution of gluon
shadowing, respectively.
}
%%%%%%%%%%%%%%%%%%%%%%%%%
 \label{r-pb-all}}
%%%%%%%%%%%%%%%%%%%%%%%%%
\end{center}
 \end{figure}
%****************************************************************
%

For a numerical solution of the Schr\"odinger
equation for the Green function 
we adopt an algorithm from ref.~\cite{n-03}.
Because available data from the E665 \cite{e665} and
NMC \cite{nmc} collaborations cover the region
of small and medium values of $Q^2 \lsim 4\,\GeV^2$
we use the KST parametrization of the dipole cross section
\cite{kst2} which is valid down to the limit
of real photoproduction. 

Performing a numerical solution
of the evolution equation for the Green function,
the imaginary part of the
LC potential (\ref{250}) contains the corresponding KST
dipole cross section as well.
The nuclear density function
$\rho_{A}({b},z)$ is taken in the 
realistic Wood-Saxon form with parameters from ref.~\cite{saxon}.
Because available
data from the E665 and NMC collaborations correspond
to very small values of $Q^2 \lsim 1\,\GeV^2$ at
small $x_{Bj}\lsim 0.004$,
the nonperturbative interaction effects
between $\bar{q}$ and $q$ are included
explicitly via the real part of the LC potential of the
form (\ref{140}).

Effects of NS are studied via $x_{Bj}$-
behavior of the ratio (\ref{340}) divided by the mass
number $A$.
Firstly we present NS for the lead target in 
Fig.~{\ref{r-pb-all}} at different fixed values of $Q^2$.
The solid and dashed curves represent the predictions 
obtained with and without contribution of GS,
respectively.

One can see that as a manifestation
of a shorter $l_c$ for higher Fock states, 
the onset of GS happens at smaller
$x_{Bj}$ than the quark shadowing. 
Fig.~\ref{r-pb-all} clearly demonstrates not very strong onset of GS in the
range of $x_{Bj}\in (0.01,0.0001)$ where the most
of available data exist. 
Besides the effects of GS are
stronger at smaller $Q^2$ because corresponding
Fock fluctuations of the photon have a larger
transverse size.

%
%****************************************************************
%************************ FIG.3 *********************************
%****************************************************************
 \begin{figure}[htb]
\includegraphics{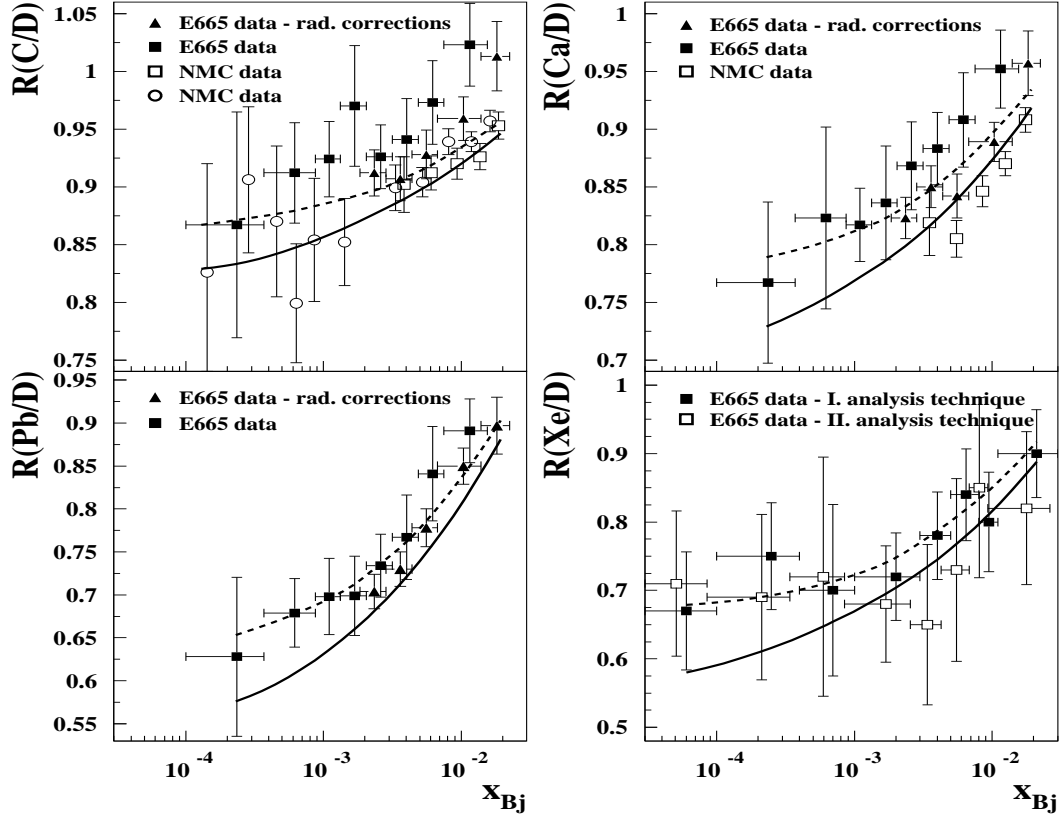}
\begin{center}
\vspace{10.80cm}
\parbox{14.0cm}
{\caption[Delta]
 {
Comparison of the model with experimental data
from the E665 \cite{e665} and NMC
\cite {nmc} collaborations. 
Calculations correspond to exact
numerical solution of the evolution equation for the Green function,
$G_{\bar{q}q}(\vec{r_2},z_2;\vec{r_1},z_1)$,
for the lowest Fock component of the photon
using KST \cite{kst2} 
parametrization of the dipole cross section and 
realistic nuclear density
function of the Woods-Saxon form \cite{saxon}. 
The solid and dashed curves
are calculated with and without contribution of gluon
shadowing, respectively.
}
%%%%%%%%%%%%%%%%%%%%%%%%%
 \label{e665-all}}
%%%%%%%%%%%%%%%%%%%%%%%%%
\end{center}
 \end{figure}
%****************************************************************
%

Saturation of NS
at low $x_{Bj}\lsim 10^{-4}$ 
at the level given by Eq.~(\ref{335})
is realized only for energy independent
dipole cross section 
(see parametrization (\ref{260})). 
However, it is not so for the
realistic energy-dependent KST parametrization,
Eq.~(\ref{kst-1}).

In Fig.~\ref{e665-all} we present a comparison of the
model predictions with data
from the E665 \cite{e665} and NMC \cite{nmc}
collaborations. Fig. shows a quite reasonable agreement
with experimental data in spite of absence of any free
parameters in the model. 
One can see that the effect of GS produces 
an additional NS,
which rises with the mass number $A$.
Fig.~\ref{e665-all} also demonstrates
a non-negligibility of GS
already in the region of $x_{Bj}\lsim 0.001\div 0.01$. 
Very large error bars 
especially at small $x_{Bj}\sim 10^{-4}$
do not allow to investigate
separately the effect of GS.
Therefore more accurate new data on NS
in DIS at small $x_{Bj}$ are very important
for the further exploratory study of nuclear
modification of the structure functions and
also GS.

%
%****************************************************************
%************************ FIG.4 *********************************
%****************************************************************
 \begin{figure}[tbh]
\includegraphics{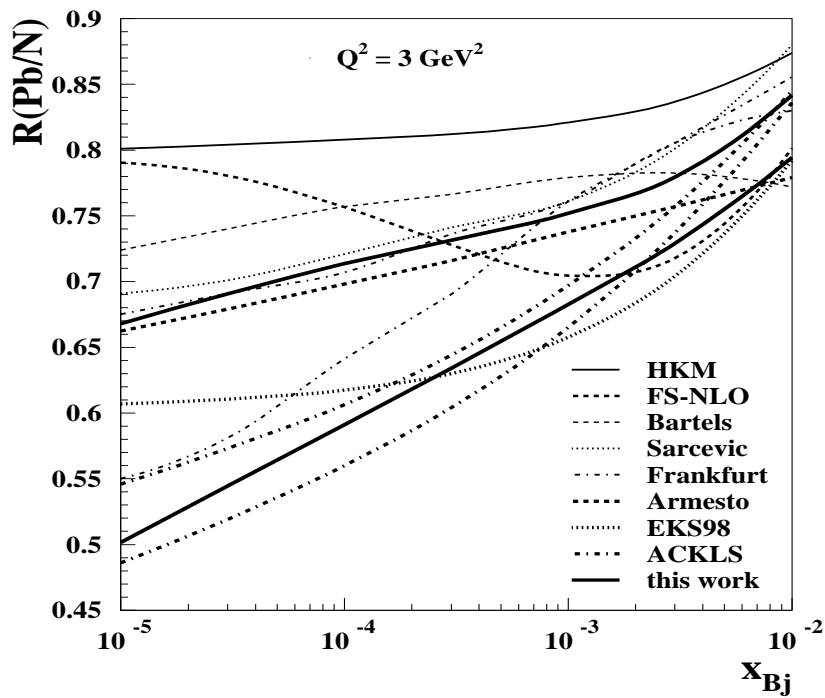}
\begin{center}
\vspace{9.20cm}
\parbox{14.0cm}
{\caption[Delta]
 {
Comparison of the model predictions for the ratio Pb/nucleon
obtained without (upper thick
solid line) and with GS (lower thick solid line) with other
models, versus $x_{Bj}$ at fixed $Q^2 = 3\,\GeV^2$.
HKM are the results from \cite{hkm}, FS-NLO from
\cite{fs-nlo}, Bartels from \cite{bartels},
Sarcevic from \cite{sarcevic}, Frankfurt from \cite{frankfurt}
($Q^2 = 4\,\GeV^2$),
Armesto from \cite{armesto}, EKS98 from \cite{eks98} and ACKLS
from \cite{ackls}.
}
%%%%%%%%%%%%%%%%%%%%%%%%%
 \label{r-pb-3-all}}
%%%%%%%%%%%%%%%%%%%%%%%%%
\end{center}
 \end{figure}
%****************************************************************
%
 
Finally,
Fig.~\ref{r-pb-3-all} represents  
a comparison of the model predictions for NS 
with the results from other models at
$Q^2 = 3\,\GeV^2$ (except the results of ref.~\cite{frankfurt}
which are at $Q^2 = 4\,\GeV^2$).
At small $x_{Bj} = 10^{-5}$ we predict quite large effects of GS
(compare upper and lower thick solid lines).
Note that the difference between models rises towards
small values of $x_{Bj}$ following from a different treatment
of various nuclear effects.

The models presented in
Fig.~\ref{r-pb-3-all} 
can be divided into several groups:

i) \emph{models where parton distribution functions (PDF's)
has been determined from data using 
Dokshitzer-Gribov-Lipatov-Altareli-Parisi
(DGLAP) 
evolution equation} \\
Here the nuclear PDF's have been
parametrized performing next to leading order (NLO) 
\cite{fs-nlo} or leading order (LO)
\cite{eks98,hkm} global
analysis of nuclear DIS and Drell-Yan data.

ii) \emph{models based on Glauber-like rescatterings}\\
Here the model in ref.~\cite{armesto} is based
on an application of a saturating ansatz for the total
$\gamma^*$-nucleon cross section in the proton.
This ansatz is then extended to the nuclear case
by its introduction in a Glauber expression.
Within the model from \cite{sarcevic} 
a Glauber ansatz provides with the initial
condition for DGLAP evolution.

iii) \emph{models based on Gribov inelastic shadowing} \\
Here in ref.~\cite{ackls} nuclear structure functions are studied
using relation with diffraction on nucleons which arises
from Gribov's reggeon calculus.
The model presented in \cite{frankfurt} 
employs again some parametrization of hard diffraction
at the scale $Q_0^2$, which gives nuclear shadowing
through Gribov's reggeon calculus similarly as in 
ref.~\cite{ackls}.
Then nuclear suppression calculated at $Q^2_0$ is
used as initial condition for DGLAP evolution.

iv) \emph{models based on high-density QCD} \\
Here the model in ref.~\cite{bartels} is based
on numerical solution of a non-linear
equation for small-$x_{Bj}$ evolution
and on application of this equation 
for the case of nuclear targets.

Fig.~\ref{r-pb-3-all} shows large differences 
in predictions of NS 
at small $x_{Bj}$ among
different models.
It gives a stimulation to obtain 
new more accurate
data on nuclear structure functions
by lepton-ion collider planned
at RHIC \cite{eRHIC}.
It can help us to discriminate among
different models.

%
%%%%%%%%%%%%%%%%%%%%%%%%%%%%%%%%%%%%%%%%%%%%%%%%%%%%%%%%%%
\section{Summary and conclusions}
\label{conclusions}
%%%%%%%%%%%%%%%%%%%%%%%%%%%%%%%%%%%%%%%%%%%%%%%%%%%%%%%%%%
%

We present a short review of the color dipole
approach based
on the LC QCD Green function formalism,
which naturally incorporates the interference 
effects of CT and CL.
Within this approach \cite{krt-98,krt-00,n-03} 
we study NS in DIS
at small Bjorken $x_{Bj}$. 

Calculations of NS corresponding
to $\bar{q}q$ component of the virtual photon 
are based on an exact numerical solution 
of the evolution
equation for the Green function.
It allows to use realistic
parametrizations of the dipole cross section 
(GBW \cite{gbw} and KST \cite{kst2}) 
and realistic nuclear density function 
of the Woods-Saxon form \cite{saxon}.

Because available data from the shadowing
region of $x_{Bj}\lsim 0.01$ coming
mostly from the E665 and NMC collaborations
cover only small and medium values
of $Q^2\lsim 4\,\GeV^2$, we prefer
KST parametrization \cite{kst2}
of the dipole cross section.
On the other hand the data obtained at
a lower part of the $x_{Bj}$- kinematic
interval correspond to very low values of
$Q^2 < 1\,\GeV^2$ (nonperturbative region).
For this reason we include explicitly
the nonperturbative interaction effects
between $\bar{q}$ and $q$ taking into
account
the real part of the LC potential 
$V_{\bar{q}q}$ (\ref{140}) in the 
time-dependent two-dimensional
Schr\"odinger equation (\ref{135})

In order to compare the realistic calculations
with data on NS, the effects of
GS are taken into account.
The same path integral technique 
\cite{kst2} is applied in this case.
GS was calculated only for the lowest Fock component, $|\bar{q}qG\ra$.
Effect of higher
Fock components containing more gluons
was essentially taken into account by eikonalization of the calculated
$R_G(x_{Bj},Q^2)$ using renormalization (\ref{1100}).
We demonstrate that the onset of GS starts to be effective 
at $x_{Bj}\sim 0.01$.
It rises
towards small $x_{Bj}$ because higher Fock components with more
gluons having shorter coherence time will contribute
to overall NS.
Such a situation is illustrated in Fig.~\ref{r-pb-all}.

Performing numerical calculations, we find that our
model is in reasonable agreement with existing
experimental data (see Fig.~\ref{e665-all}).
Large error bars and incompatibility
of the experimental results from the E665 and NMC 
collaborations do not allow 
to study separately the effect of GS.
Therefore more accurate new data on NS
in DIS off nuclei at still smaller $x_{Bj}\lsim 10^{-5}$ are very important
for the further exploratory study of GS effects.

Comparison among various models shows large differences
for the Pb/nucleon ratio of the structure functions at
$x_{Bj} = 10^{-5}$ and $Q^2 = 3\,\GeV^2$ (see Fig.~\ref{r-pb-3-all}).
It has a large impact on the calculation of high-$p_T$ particle
spectra
in  nuclear collisions at RHIC and LHC.
Such large differences at small $x_{Bj}$ among
different models should be testable by the 
new more precise
data on nuclear structure functions   
which can
be obtained by lepton-ion collider planned
at RHIC \cite{eRHIC}.

\medskip

\noindent
 {\bf Acknowledgments}:
This work has
been supported in part by the Slovak Funding Agency, Grant No. 2/7058/27
and by grant VZ MSM 6840770039, and LC 07048 (Czech Republic).

%%%%%%%%%%%%%%%%%%%%%%%%%%%%%%%%%%%%%%%%%%%%%%%%%%%%%%%%%%
 \def\appendix{\par
 \setcounter{section}{0} \setcounter{subsection}{0}
 \def\thesection{Appendix \Alph{section}}
\def\thesubsection{\Alph{section}.\arabic{subsection}}
\def\theequation{\Alph{section}.\arabic{equation}}
\setcounter{equation}{0}}
%%%%%%%%%%%%%%%%%%%%%%%%%%%%%%%%%%%%%%%%%%%%%%%%%%%%%%%%%%

\end{document}